# Symmetry breaking in covalent chiral bond H$_2$, according to accurate vibrational levels from Kratzer bond theory

*G. Van Hooydonk, Ghent University, Faculty of Sciences, Krijgslaan 281 S30, B-9000 Belgium*
*Y.P. Varshni, University of Ottawa, Department of Physics, Ottawa, Canada*

Abstract. *Symmetry breaking in H$_2$, quantified with Kratzer bond theory, leads to vibrational levels with errors of only 0,00008 %. For quanta, 0,0011 % errors are smaller than with any ab initio QM method. Chiral behavior of covalent bond H$_2$ implies bonding between left- and right-handed atoms H$_R$ and H$_L$ or between hydrogen H and antihydrogen $\underline{H}$. This generic H$_2$ asymmetry is given away by a Hund-type Mexican hat curve, invisible in QM.*

## I. Introduction

Symmetry breaking (SB) at low energy in neutral, stable, small quantum systems is important. H and H$_2$, the simplest but most abundant systems in the Universe [1], being prototypical for atomic and molecular spectroscopy [2], are therefore prototypical for SB at the eV-level. Unfortunately, bound state H QED [3] and H$_2$ QM theories [4-6] are complex: QM relies on parameters and hundreds of terms in the wave function to get at observed H$_2$ levels [7]. The lack of an analytical potential energy function (PEF)[8] for H$_2$ in [4-6] is also unfortunate: this PEF (i) may disclose the long sought for simple low parameter universal function (UF) behind all shape-invariant potential energy curves (PECs) [2,8-11] and (ii) may, eventually, disclose SB in H$_2$. Hence, a simpler H$_2$ bond theory is of interest for SB but only if it is more accurate than QM, which is problematic. In fact, QM [6] may be the most precise H$_2$ theory, consistent with observed data [12], it overlooks SB. A recent Kratzer H$_2$ bond theory [13] gave errors of 3,4 cm$^{-1}$, comparable with those of 3,2 cm$^{-1}$ in earlier QM [4]. Whereas their large errors vanish with non-adiabatic corrections [5-6], the 3,4 cm$^{-1}$ errors in [13] vanish with an equally simple, parameter free chiral Kratzer theory, as we show here. Being more precise for H$_2$ than any ab initio QM theory available, this theory for the chemical bond deals analytically with symmetry breaking in H$_2$, which is difficult, if not impossible with ab initio QM.

## II. Dunham and Kratzer oscillators for (too) symmetrical H$_2$

The standard 4 particle (a$^-$,A$^+$; b$^-$,B$^+$) Hamiltonian **H** or classical energy E for H$_2$

$$E=\mathbf{H}=\tfrac{1}{2}p_a^2/m_a+\tfrac{1}{2}p_b^2/m_b+\tfrac{1}{2}p_A^2/m_A+\tfrac{1}{2}p_B^2/m_B-e^2/r_{aA}-e^2/r_{bB}-e^2/r_{bA}e^2/r_{aB}+e^2/r_{ab}+e^2/r_{AB} \quad (1a)$$

where all symbols have their usual meaning, led to vibrational energies [13]

$$E_{vib}=\Delta \mathbf{H}=(E-2E_H)\approx f^2\hbar^2/(mr_{AB}^2)\pm A_r e^2/r_{AB}=B/r^2\pm A_r e^2/r \quad (1b)$$

With the B-term positive, the $A_r$-term can only be negative for H$_2$ to be stable. Scaling by $\Delta\mathbf{H}_0 = -\tfrac{1}{2}A_r e^2/r_0$ gives Kratzer oscillator K(r$_0$/r) in variable r$_0$/r [14]

$$K(r_0/r)+1=\Delta\mathbf{H}/(-\tfrac{1}{2}A_r e^2/r_0)+1=(r_0/r)^2-2r_0/r+1=(1-r_0/r)^2 \quad (1c)$$

analytically and conceptually different from JWKB and Dunham potentials in variable r/r$_0$ [15]. Dunham's V$_D$(r/r$_0$) and Kratzer's V$_K$(r$_0$/r) potentials are respectively

$$V_D(r) = a_0(1-d)^2=(1-r/r_0)^2 \quad (1d)$$

$$V_K(1/r) = a_0(1-1/d)^2=a_0(1-r_0/r)^2=(a_0/r^2)V_D(r) \quad (1e)$$

With $U(r)=\Sigma a_n(r-r_0)^n$, Dunham's oscillator (1d) relates to the JWKB-approximation. Kratzer's (1e) refers naturally to turning points $(e^2/r_0)(r_0/r_- - r_0/r_+)$ in PECs, as disclosed by RKR-methods [16-18]. In [13], deviations from $r_0$ in $r=r_0+\Delta$, $r/r_0=1+\Delta/r_0=1+\delta$ are quantized with vibrational number v using $\delta=qv$. Equally distributed in perfectly symmetric $H_2$, these deviations generate [13]

$$\delta=r/r_0-1=\Delta/r_0=\tfrac{1}{2}\Delta/r_0-(-\tfrac{1}{2}\Delta/r_0)=\tfrac{1}{2}qv-(-\tfrac{1}{2}qv)=qv \qquad (1f)$$

$$q=\omega_e/a_0=2\omega_e/(e^2/r_0)=\omega_e/(\tfrac{1}{2}k_e r_0^2)=\omega_e/(\tfrac{1}{2}D_{ion})=4410{,}172/78844{,}913=0{,}05591 \qquad (1g)$$

where $\omega_e$ is the fundamental frequency, $a_0$ the first order Dunham coefficient, $D_{ion}=e^2/r_0$ the ionic bond energy and force constant $k_e=e^2/r_0^3$ [13]. The advantage over [4-6] is that $H_2$ characteristics in (1g) all derive directly from atom mass $m_H$ and its classical radius, defined as [13]

$$r_H=[3m_H/(4\pi\Gamma_H)]=0{,}7365 \cdot 10^{-8} \text{ cm} \qquad (1h)$$

for density $\Gamma_H=1$ [13,19]. Kratzer's potential (1c) in inverse $r_0/r$ or

$$\delta_K=r_0/r_1-r_0/r_2=1/(1-\tfrac{1}{2}\Delta/r_0)-1/(1+\tfrac{1}{2}\Delta/r_0)=qv/(1-\tfrac{1}{4}q^2v^2) \qquad (1i)$$

led to level results for $H_2$, 30 times more accurate than Dunham's (1f) [13].

Since the term in $A_r$ in (1b) must be <0, Coulomb law for an ion pair with reduced mass $\mu_{HH}=\tfrac{1}{2}m_H$ gives $E_{HH}=\tfrac{1}{2}\mu_{HH}v^2-e^2/r$. Its first derivative $d/dr$ gives classical radial equilibrium condition

$$\tfrac{1}{2}m_H v^2 r=e^2 \qquad (1j)$$

Using $m_H$ and $r_H$ in (1h), velocity v in (1j) leads to a $H_2$ fundamental vibrational frequency [13]

$$\omega_H=4410{,}172 \text{ cm}^{-1} \qquad (1k)$$

close to the value 4401 cm$^{-1}$ in [7,20]. Since 3,4 cm$^{-1}$ errors for symmetric $H_2$ do not comply with spectroscopic accuracy [13], we now consider a less symmetrical, chiral $H_2$ model.

## III. Symmetry-breaking from achiral to chiral $H_2$

Following the chemist's symmetry view $H_2=2H=HH$, bisecting $H_2$ line segment $L=r_0$

$$H_2=\{H_L,C,H_R\} \sim L(0,+\tfrac{1}{2},+1) \text{ or } L'(-\tfrac{1}{2}, 0, +\tfrac{1}{2}) \qquad (2a)$$

implies a bond symmetry $S_0$, quantified by the ratio of equal parts (proportions), i.e. $S_0=\tfrac{1}{2}/\tfrac{1}{2}=1$, valid for any $r_{AB}>$ or $<r_0$. Deviations $\delta<0$ or $\delta>0$ from $\tfrac{1}{2}r_0=r_H$ do not alter $S_0$, since

$$S_0=\tfrac{1}{2}(1\pm\delta)/[\tfrac{1}{2}(1\pm\delta)]=\tfrac{1}{2}/\tfrac{1}{2}=1 \qquad (2b)$$

remains valid, however large $\delta$. $H_L$ at the left and $H_R$ at the right of the center call for back-front or mirror-symmetry in $H_2$. The frame is left-handed for $H_L$ and right-handed for $H_R$ (or vice versa) but $S_0$ typifies a too symmetrical, achiral unit $H_2$, although using $H_L H_R$; $H_R H_L$ is superfluous, if $H_L=H_R$ [10]. As in QM, there is no need for less symmetrical or chiral $H_2$, for which $H_L \neq H_R$ and $S \neq 1$. Theoretically, sign-conjugated deviations $\gamma$ from achiral part $p_a=\tfrac{1}{2}$ always give generic unequal parts

$$p_\gamma=p_{R,L}=p_\pm=\tfrac{1}{2}(1\pm\gamma)=p_a(1\pm\gamma)=p_L+p_R \qquad (2c)$$

The ratio of chiral parts $p_L/p_R$ returns an intrinsic, generic $H_2$ left-right asymmetry (chirality[1,2,3])

---

[1] Difference $\gamma$ is a *continuous chirality measure* (CCM) [21].
[2] Left and right are formalized with Dirac's $\gamma^5$ [22]. Dimensionless left and right properties P are $P_L=\tfrac{1}{2}p(1-\gamma_L)$ and $P_R=\tfrac{1}{2}p(1+\gamma_R)$, with $|\gamma_L|=|\gamma_R|$, implying that *centers of chiral systems are not exactly in the middle*.

$$S_C = (1-\gamma)/(1+\gamma) \tag{2d}$$

equal to $S_0$ only if asymmetry effect $\gamma=0$. Complementarity $p_R=1-p_L$ or $x_2=1-x_1$ gives

$$1=x_1+x_2=p_L+p_R=x_1+(1-x_1)=p_L+(1-p_L)=p_R+p_L=\tfrac{1}{2}(1+\gamma)+\tfrac{1}{2}(1-\gamma)=1 \tag{2e}$$

For parts not equal to ½, an axial view is needed. Parallel ⌊⌋, ⌈⌉ or anti-parallel ⌊⌉, ⌈⌋ states in 3D $H_2$ view (1h) use all intra- and inter-atomic separations $r_{Aa}$, $r_{Bb}$, $r_{AB}$, $r_{ab}$, $r_{Ab}$ and $r_{Ba}$ in (1a). For $X_2$ bonds, reduced axial and radial parts are $|\tfrac{1}{2}|$ and $|1|$. At $r_0=2r_H$, hypotenuse $h=r_{Ab}=r_{Ba}$ for axial $H_2$ states $e^2/r_{Ab}$ and $e^2/r_{Ba}$ in (1a) in reduced form is equal to

$$h/(2r_H)=h' = (1/2r_H)\sqrt{(4r^2_H+r^2_H)}=\tfrac{1}{2}\sqrt{5} \tag{2f}$$

invariantly giving away the square root of 5 by definition. Upon bisection, its 2 equal parts

$$\tfrac{1}{2}h'=\tfrac{1}{4}\sqrt{5}=0{,}55901699>\tfrac{1}{2} \tag{2g}$$

are larger than achiral value ½ by exactly 0,059026994 or 1/0,059026994 (≈17), commensurate with (1g). Classically, axial states (2f) use Euclidean division [19] (see Section V) and may well lead in a generic way to $\gamma$-effects related to $\sqrt{5}$, overlooked thus far in all $H_2$ theories, including QM.

**IV. Formal Kratzer bond theory for symmetry breaking in chiral $H_2$**

In theory, the effect of non-zero $\gamma$ on the $H_2$ structure is easily quantified with oscillators (1g)-(1h).

(a) Linear Dunham variable $\delta$ (1f) is $\gamma$-invariant: achiral and chiral cases are degenerate

$$\text{achiral: } qv=+\tfrac{1}{2}qv+\tfrac{1}{2}qv=\tfrac{1}{2}qv-(-\tfrac{1}{2}qv)=qv \tag{3a}$$

$$\text{chiral: } qv=+\tfrac{1}{2}(1+\gamma)qv+\tfrac{1}{2}(1-\gamma)qv=\tfrac{1}{2}qv+\tfrac{1}{2}qv=\tfrac{1}{2}qv-(-\tfrac{1}{2}qv)=qv \tag{3b}$$

(b) With inverse Kratzer variable $\delta_K$ (1i), this $\gamma$-degeneracy is lifted since

$$\text{achiral: } 1/(1-\tfrac{1}{2}qv) - 1/(1+\tfrac{1}{2}qv) = qv/(1-\tfrac{1}{4}q^2v^2) \tag{3c}$$

$$\text{chiral: } 1/[1-\tfrac{1}{2}(1+\gamma)qv] - 1/[1+\tfrac{1}{2}(1-\gamma)qv] = qv/[1-\gamma qv-\tfrac{1}{4}(1-\gamma^2)q^2v^2] \tag{3d}$$

The formal effect of non-zero $\gamma$ (symmetry breaking) in $H_2$ in a Kratzer variable is

$$qv/[(1-\tfrac{1}{4}q^2v^2)-\gamma qv+\tfrac{1}{4}\gamma^2 q^2v^2] = [qv/(1-\tfrac{1}{4}q^2v^2)]/[1-\gamma qv(1-\tfrac{1}{4}\gamma qv)/(1-\tfrac{1}{4}q^2v^2)] \tag{3e}$$

The ratio of achiral (3b) and (3c) gives harmonic mean $(1-\tfrac{1}{4}q^2v^2)=(1+\tfrac{1}{2}qv)(1-\tfrac{1}{2}qv)$, the reason why Kratzer's oscillator outperforms Dunham's by a factor 30 [9,13]. Since harmony improves $H_2$ results [13], other harmonies, including those with $\gamma$ as in (3e), must be inventoried (see Section VI).

**V. Euclidean $H_2$ symmetry**

Axial states in Section III are either parallel or anti-parallel. Parallel boat structure (4a)

$$\text{(L)} \quad \begin{array}{c} r_H \\ \text{▱} \\ 2r_H \end{array} \quad \text{(R)} \quad \text{and} \quad \text{(L)} \quad \begin{array}{c} \text{▱} \quad r_H \\ 2r_H \end{array} \quad \text{(R)} \tag{4a}$$

contains 2 rectangular triangles, one left-, the other right-handed (or vice versa). Since these cannot coincide without leaving the paper plane, chirality applies (mirror, perpendicular to the paper).

---

[3] In Heitler-London theory [24], permutation, achieved with two-center functions $\psi_{AB}$ and $\psi_{BA}$, leads to exchange forces, responsible for bonding, whereby chiral behavior is not considered.

Chair structures are achiral, not chiral, unless L- and R-parts are unequal (S≠1). Unlike (4a), the 2 triangles in a chair can be made to coincide by in-plane rotation. With different sizes, a perspective will displace them in front and back of the mirror in the paper plane.

In either case, Euclidean division of $AC=r_{Ab}=r_{Ba}=AB+AC=a+b=a(1+k)$ and number $k=b/a$, gives

$$AB/BC=BC/AC; \quad a/b=b/(a+b) \text{ or } 1/k=k/(1+k) \qquad (4b)$$

This brings in $k^2-k-1=0$ and solutions $k=\frac{1}{2}(1\pm\sqrt{5})$. Golden ratio $k=b/a$ obeys phi-numbers[4]

$$\varphi=1/\Phi=\frac{1}{2}(1+\sqrt{5}) \qquad (4c)$$

in line with (2f)-(2h) [19]. Strangely enough, (4c) is not the only solution possible.

---

Table 1 Phidias-Euclid and Dirac schemes for complementary chiral parts in $H_2$

|  | Phidias-Euclid | Complementarity | Dirac I | Dirac II[a] |
|---|---|---|---|---|
| Left part $x_L$ | 1 | x | $\frac{1}{2}(1-\gamma)$ | $\frac{1}{2}-\gamma'$ |
| Right part $x_R$ | k | 1-x | $\frac{1}{2}(1+\gamma)$ | $\frac{1}{2}+\gamma'$ |
| Unit | 1+k | 1 | 1 | 1 |
| Ratio's | $1/k=k/(1+k)$ | $x/(1-x)=1-x$ | $(1-\gamma)/(1+\gamma)=\frac{1}{2}(1+\gamma)$ | $(\frac{1}{2}-\gamma')/(\frac{1}{2}+\gamma')=\frac{1}{2}+\gamma'$ |
| Quadratic | $k^2-k-1=0$ | $x^2-3x+1=0$ | $\gamma^2+4\gamma-1=0$ | $\gamma'^2+2\gamma'-\frac{1}{4}=0$ |
| Solutions[b] | $k=\varphi=\frac{1}{2}(1\pm\sqrt{5})$ | $x=(3/2)(1\pm\sqrt{5}/3)$ | $\gamma=-2(1\pm\frac{1}{2}\sqrt{5})$ | $\gamma'=-(1\pm\frac{1}{2}\sqrt{5})$ |
| Values[c] | +1,618; -0,618 | +2,618; +0,382 | -4,236; +0,236 | -0,118; +2,118 |
| With inverse | $k=1+1/k$ | $x=3-1/x$ | $\gamma=1/\gamma-4$ | $\gamma'=1/(4\gamma)-2$ |

---

a) Dirac I solution $\frac{1}{2}(1\pm\gamma)$ transforms in $\frac{1}{2}\pm\frac{1}{2}\gamma=\frac{1}{2}\pm\gamma'$, with handedness $\gamma=2\gamma'$ (see solutions for $\gamma$ and $\gamma'$).
b) Interchanging $x_R$ and $x_L$ gives different quadratics and solutions: $k^2+k-1$, $x^2+x-1=0$ and $\gamma^2-4\gamma-1=0$. A permutation of parts leads to 4 rather than 2 solutions of type $\pm a(1\pm b\sqrt{5})$, not given in the Table.
c) Only 3 decimals given, based on $\sqrt{5}=2,236067978…\approx 2,236$

Alternatives, all containing $\sqrt{5}=2\varphi-1$, are in Table 1. The 4 different scale factors for units are ½, 1, 3/2 and 2; the 3 different coefficients for $\sqrt{5}$ are 1, 1/3 and ½. The last 3 Columns apply for unit 1; Phidias-Euclid recipe in Column 2 treats one part as if it were the unit. Choices are difficult by relations between linear k, x, $\gamma$ and inverse $1/k$, $1/x$, $1/\gamma$ (see last row). Table 1 does not single out a ***best*** solution. Rather than solving the wave equation for Hamiltonian (1a), we test all solutions possible by plugging them in (1l)-(1m) and looking at the results obtained.

**VI. Ionic chiral Kratzer bond theory for $H_2$ with left-right asymmetric $H_L$ and $H_R$**

Of all possible combinations in Table 1, only parts $x_{RL}$, based on $\sqrt{5}$ as in (2f) and (4c) and equal to

$$x_{RL}=x_{\pm}=p(1\pm\gamma)=(2/3)(1\pm\frac{1}{2}/\varphi)=(2/3)(1\pm\frac{1}{2}\Phi) \qquad (5a)$$

invariantly related to Euclid's golden number

$$\Phi=1/\varphi=2/(1+\sqrt{5})=\frac{1}{2}(\sqrt{5}-1)=0,618033989…. \qquad (5b)$$

reproduce $H_2$ levels within greater precision than QM (see below). Plugging (5a) in (3f) gives

$$\delta_K=r_0(1/r_a-1/r_b)=1/[1-(1+\frac{1}{2}\Phi)qv/3]-1/[1+(1-\frac{1}{2}\Phi)qv/3]$$
$$=(2/3)qv/[1-\Phi qv/3-q^2v^2(1-\frac{1}{4}\Phi^2)/9]=(2/3)qv/[1-0,206011qv-0,100501q^2v^2] \quad (5c)$$

whereby internal $H_2$ asymmetry is assessed with Euclid's recipe involving axial states.

---

[4] Phi-numbers appear in mathematics (Fibonacci series…), physics, chemistry, biology, architecture, arts [25].

As in [13], multiplying (5c) with 1,5 to correct for Euclidean factor 2/3 in (5a) and (5c) returns

$\delta'=1,5\delta_K=qv/(1-0,206011qv-0,100501q^2v^2)$  (5d)

the parameter free chiral Kratzer variable, we use below for fitting $H_2$ levels [7]. The chiral version $v_\gamma$ of the conventional vibrational quantum number v is an effective quantum number

$v_\gamma=\delta''=\delta'/q=1,5\delta_K/q=v/(1-0,206011qv-0,100501q^2v^2)$  (5e)

close to $v/[1-(qv/5)(1+½qv)]$. Fitting with (5e) gives coefficients smaller by q, see (6b) below.

**VII. Results**

*VII.1 $H_2$ levels and bond energy $D_e$*

Fitting the 14 $H_2$ levels [7] with (5d) using a quartic through the origin gives

$E_\delta=-4864,602868\delta^4+18697,327977\delta^3-54425,081623\delta^2+76533,833034\delta$ cm$^{-1}$  (6a)

with goodness of fit $R^2=0,999999999997$. As in [13], the term in $\delta$ has the correct value, close to $a_0$ in (1g). A fit with (5e) leads to the more familiar quartic in v (similar to that in (v+½) [13])

$E_v=-0,047618v_\gamma^4+3,272089v_\gamma^3-170,279673v_\gamma^2+4280,902374v_\gamma$  (6b)

The main advantage of (6a) [13] is that the $H_2$ bond energy $D_e$ is given by the intercept, appearing when level energies are plotted versus complementary variable

$x=1-b\delta=1-1,5.0,927629981107\delta=1-1,3914449972\delta$  (6c)

for this makes the linear Coulomb term in $\delta$ in (6a) vanish exactly. The same factor 1,3914 also appears for the H spectrum as $r_H/r_B=1,3915…$, where $r_B$ is the Bohr length, and close to $9\varphi/4$ [19]. For achiral $H_2$, $D_e$ is 36146,44 cm$^{-1}$ [13]. For chiral $H_2$ (6c), $D_e$ appears in a closed form quartic

$E_x=-6569,703251x^4+2855,209522x^3-32395,749724x^2+36110,244712$ cm$^{-1}$  (6d)

$=-[6569,703252x^4-2855,209522x^3+310,220306x^2]-32085,529418x^2+D_e$ cm$^{-1}$  (6e)

mathematically equivalent to and as precise as (6a). Fig. 1 illustrates the effect on levels of adjusted and complementary variables (6c). The quartic in (6d) exposes the asymmetrical chiral nature of $H_2$, although this contribution is relatively small. Fig. 2 shows the Hund-type $H_2$ Mexican hat curve

$(D_e-E_x)-32085,529418x^2=6569,703252x^4-2855,209522x^3+310,220306x^2$  (6f)

It exposes new critical points, due to left-right asymmetric, chiral $H_2$. Fig. 2 also shows the curve for terms in $x^3$ and $x^4$ in (6d). Fig. 3 zooms in on these new critical points for $H_2$, given away by its vibrational spectrum but invisible in QM, and which typify symmetry breaking in $H_2$.

*VII.2 Precision of parameter free chiral Kratzer bond theory: comparison with ab initio QM*

Level errors of 0,015 cm$^{-1}$ give a precision of $8,6.10^{-7}$ %, see Table 2. With 0,05 cm$^{-1}$ errors for Dabrowski data [7], the constraint of spectroscopic accuracy is met. 4$^{th}$ and 6$^{th}$ order fits with qv give errors of 7,15 cm$^{-1}$ and 0,24 cm$^{-1}$, 475 and 17 times larger than a 4$^{th}$ order fit with (5d). Table 3 for $\Delta G(v+½)$ includes errors of all ab initio QM methods available [4-6,26-28]. Error ratios (%) vary from 40,3 for 1975 QM [4] to 1,8 for 1995 QM with many correction terms [6]. Recent QM methods [27,28] are less precise (see last row).

**VIII. Discussion**

(i) The centuries old problem [2, 9-11] with ionic and covalent energies $D_{ion}$ and $D_e$ is solved. With (6c)-(6d), $D_e$ is generated analytically by Coulomb's ionic bond energy $D_{ion}$, securing the $H_2$ bond is stable [13]. In an effortless way, with an ionic Coulomb view and with (6a)-(6c), $D_e$ amounts to

$$D_e = 36110{,}244711 \text{ cm}^{-1} \qquad (7a)$$

Although slightly lower than 36118,3 cm$^{-1}$ in [20], the deviation of 8 cm$^{-1}$ is only 0,022 %. A similar difference appears between $\omega_e$=4410,1722 cm$^{-1}$ in (1g) and 4401,213 cm$^{-1}$ in [20]. This result is also important for the distinction between $D_{ion}$ and $D_e$ as a scaling aid the molecular constants and in the search for the universal function (UF) [2,8-11] (see Introduction).

(ii) The unprecedented precision in this work derives from only one parameter free variable (5c) and only 3 terms in $x^2$, $x^3$ and $x^4$ in (5g). This analysis outperforms QM [4-6, 26-28], although all these QM methods are highly parameterized and use hundreds of terms in the $H_2$ wave function.

(iii) Of all QM methods in Table 3, Wolniewicz's method [6] may be the best [12], it is still 2 times less precise than ours. Wolniewicz used relativistic, adiabatic and non-adiabatic corrections with ab initio QM in a BO-approximation [6]. These corrections, as well as QM itself, are all avoided in a simple chiral Kratzer approach, which, nevertheless, remains the more precise (see Table 3).

(iv) Errors for $H_2$ quanta in Table 3 are of the same order as the standard H Lamb shift. Hence, our results call for new determination of $H_2$ levels with a precision of 0,001 cm$^{-1}$ or better. These may settle problems with $P_{1/2}$ or $S_{1/2}$ states for the $H_2$ ground state and confirm the quality of our results.

(v) Although simple first principles chiral Kratzer $H_2$ bond theory uses only hydrogen mass $m_H$ as input, new critical points emerge, invisible in and never exposed with ab initio QM (see Fig. 2-3).

(vi) Whereas the potential in the JWKB-approximation starts off with linear $k(r_1-r_2)$ as in a Dunham expansion, it is evident from all Coulomb terms in (1a) and from RKR-procedures that a potential in inverse r or 1/r, say $e^2/r_1 - e^2/r_2 = (e^2/r_0)(r_0/r_1 - r_0/r_2)$ seems superior.

(vii) A chiral $H_2$ bond must be interpreted with CP [13]. Reminding (1c) and the $A_r$-term, constant $A_r$ implies that $H_2$ geometry is fixed. This excludes coordinate dependent P-effects but points to intra-atomic charge inversion C, for only a term in $A_r<0$ can make $H_2$ stable [10,23]. Then, our results provide with signatures for natural antihydrogen- or $\underline{H}$-states [10,23,29,30]. The common sense idea [10] that $H_2$ consists of $H_L H_R$ and $H_R H_L$ (or of H$\underline{H}$ and $\underline{H}$H) is given away by Hund-type Mexican hat curves for $H_2$ (Fig. 2-3). To make sense [29,30], also the H line spectrum must exhibit left-right asymmetry, point to $H_R$- and $H_L$-states or to H- and $\underline{H}$-states through the intermediary of a similar H Mexican hat curve for natural atom H, which is exactly what we observed [29,30].

(vii) The rigor of ab initio QM, often contra productive and inconclusive, can be avoided with less rigorous density functional theory (DFT) [10], seemingly in line with density Γ in (1h). Coefficient 1,391.. in (5f) for $H_2$ bond densities also appears identically for H atom density [19].

(viii) Using (5a) and (5b), generic asymmetry $S_C$ in (2d) for $H_2$, is now related quantitatively to

$$S_C \sim (1-\tfrac{1}{2}\Phi)/(1+\tfrac{1}{2}\Phi) = 0{,}690983/1{,}309017 = 0{,}527864 \qquad (7b)$$

as given away by the $H_2$ vibrational spectrum [7], the backbone of the $H_2$ PEC.

**IX. Conclusion**

Conceptually simple ionic Kratzer chiral bond theory is accurate for the prototypical and simplest quantum oscillator in nature: covalent bond $H_2$. *Asymmetrical, chiral* $H_2$ binds hydrogen (H-state) to antihydrogen (H̲-state). Wave equation and wave functions are not needed, since the first principles of old quantum theory suffice [13]. This simpler theory proves more accurate than any ab initio QM $H_2$ theory available. Unlike QM or QED, low energy symmetry breaking or left-right asymmetry in both H and $H_2$, eventually leads to even more accurate, analytical solutions than hitherto believed.

Table 2. Experimental [7] and theoretical vibrational energy levels of $H_2$ (cm$^{-1}$)

| v | $E_{v,0}$ [7] | This work | Difference |
|---|---|---|---|
| 0 | 0,00 | 0,000 | 0,000 |
| 1 | 4161,14 | 4161,143 | -0,003 |
| 2 | 8086,93 | 8086,943 | -0,013 |
| 3 | 11782,36 | 11782,321 | 0,039 |
| 4 | 15250,31 | 15250,317 | -0,007 |
| 5 | 18491,92 | 18491,917 | 0,003 |
| 6 | 21505,78 | 21505,799 | -0,019 |
| 7 | 24287,91 | 24287,950 | -0,040 |
| 8 | 26831,16 | 26831,128 | 0,032 |
| 9 | 29124,09 | 29124,081 | 0,009 |
| 10 | 31150,47 | 31150,442 | 0,028 |
| 11 | 32887,13 | 32887,155 | -0,025 |
| 12 | 34302,20 | 34302,206 | -0,006 |
| 13 | 35351,36 | 35351,358 | 0,002 |
| 14 | 35973,38 | 35973,377 | 0,003 |

average difference 0,0151 cm$^{-1}$ (8,6.10$^{-7}$ %)

Table 3. Experimental and theoretical quanta for $H_2$ and differences ε (Exp-Theo in cm$^{-1}$)

| | Quanta $\Delta G(v+½)$ | | Differences [a] in this work and in 7 QM studies from 1975 to 2008 as referenced | | | | | | |
|---|---|---|---|---|---|---|---|---|---|
| v | Exp [7] | *This work* | *This work* | 1975[4] | 1983[5] | 1993[6] | 1995[6] | >1995[26] | 2006[27] | 2008[28] |
| 0 | 4161,14 | 4161,143 | -0,002 | -0,94 | -0,04 | -0,027 | -0,027 | -0,023 | -0,024 | -0,0241 |
| 1 | 3925,79 | 3925,800 | -0,010 | -0,88 | -0,07 | -0,052 | -0,046 | -0,047 | -0,049 | -0,0484 |
| 2 | 3695,43 | 3695,379 | 0,051 | -0,74 | 0,01 | 0,029 | 0,041 | 0,038 | 0,035 | 0,0354 |
| 3 | 3467,95 | 3467,996 | -0,046 | -0,69 | -0,07 | -0,037 | -0,026 | -0,033 | -0,036 | -0,0357 |
| 4 | 3241,61 | 3241,600 | 0,010 | -0,50 | 0,02 | 0,036 | 0,046 | 0,033 | 0,029 | 0,0301 |
| 5 | 3013,86 | 3013,881 | -0,021 | -0,48 | -0,02 | -0,001 | 0,009 | -0,009 | -0,012 | -0,0116 |
| 6 | 2782,13 | 2782,152 | -0,022 | -0,38 | -0,02 | -0,024 | -0,006 | -0,031 | -0,036 | -0,0340 |
| 7 | 2543,25 | 2543,178 | 0,072 | -0,20 | 0,08 | 0,043 | 0,075 | 0,041 | 0,037 | 0,0388 |
| 8 | 2292,93 | 2292,953 | -0,023 | -0,12 | -0,03 | -0,067 | -0,020 | -0,063 | -0,067 | -0,0644 |
| 9 | 2026,38 | 2026,361 | 0,019 | 0,15 | -0,05 | -0,030 | 0,029 | -0,026 | -0,028 | -0,0258 |
| 10 | 1736,66 | 1736,712 | -0,052 | 0,27 | -0,15 | -0,108 | -0,047 | -0,116 | -0,118 | -0,1156 |
| 11 | 1415,07 | 1415,052 | 0,018 | 0,69 | 0,08 | -0,043 | -0,006 | -0,093 | -0,092 | -0,0906 |
| 12 | 1049,16 | 1049,152 | 0,008 | 1,11 | -0,06 | 0,038 | 0,021 | -0,090 | -0,087 | 0,0444 |
| 13 | 622,02 | 622,019 | 0,001 | 1,70 | 0,30 | 0,164 | 0,064 | -0,078 | -0,068 | -0,2021 |
| Error in cm$^{-1}$ | | | 0,025 | 0,632 | 0,071 | 0,050 | 0,033 | 0,036 | 0,051 | 0,0572 |
| Error in % | | | 0,0011 | 0,0424 | 0,0059 | 0,0037 | 0,0019 | 0,0033 | 0,0032 | 0,0044 |
| Ratio % with this work | | | 1 | 40,3 | 5,6 | 3,5 | 1,8 | 3,1 | 3,0 | 4,2 |

a) all decimals as given in published data

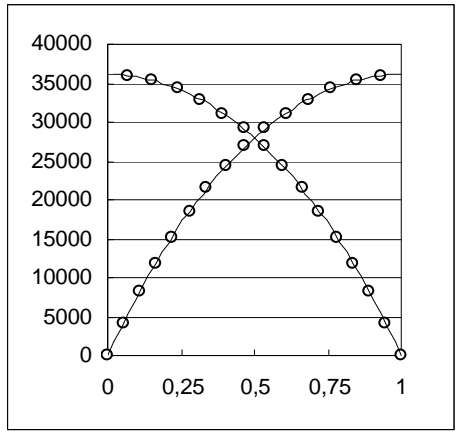

Fig. 1 Levels $E_{v,0}$ versus Euclidean $b\delta$ (left to right) and complementary $x=1-b\delta$ (right to left), with $D_e$ as natural intercept

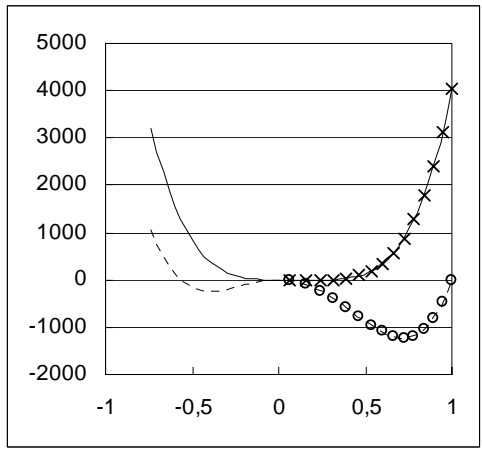

Fig. 2 H$_2$ Mexican hat curves: eqn (6f) (full line x) and $D_e(1-x^2)-E_{v,0}$ (dashed line o), both quartics extrapolated to the left

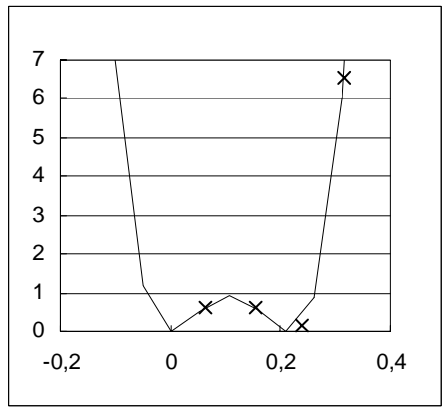

Fig. 3 Zooming in on the lower part of the H$_2$ Mexican hat curve, eqn. (6f)